\documentstyle[12pt]{article}
\begin{document}
\centerline{\large\bf CP violation effect in long-baseline neutrino oscillation }
\vskip 0.6truecm
\centerline{\large\bf in the four-neutrino model}
\baselineskip=8truemm
\vskip 2.5truecm
\centerline{Toshihiko Hattori$,^{a),}$\footnote{e-mail: 
hattori@ias.tokushima-u.ac.jp} \ Tsutom Hasuike$,^{b),}$\footnote{e-mail: 
hasuike@anan-nct.ac.jp} \ and \ Seiichi Wakaizumi$ ^{c),}$\footnote{e-mail: 
wakaizum@medsci.tokushima-u.ac.jp}}
\vskip 0.6truecm
\centerline{\it $ ^{a)}$Institute of Theoretical Physics, University of Tokushima,
Tokushima 770-8502, Japan}
\centerline{\it $ ^{b)}$Department of Physics, Anan College of Technology,
Anan 774-0017, Japan}
\centerline{\it $ ^{c)}$School of Medical Sciences, University of Tokushima,
Tokushima 770-8509, Japan}
\vskip 2.5truecm
\centerline{\bf Abstract}
\vskip 0.7truecm

We investigate CP-violation effect in the long-baseline neutrino oscillation 
in the four-neutrino model with mass scheme of the two nearly degenerate pairs 
separated with the order of 1 eV, by using the data from the solar neutrino deficit, 
the atmospheric neutrino anomaly and the LSND experiments along with the 
other accelerator and reactor experiments. By use of the most general 
parametrization of the mixing matrix with six angles and six phases, we show 
that the genuine CP-violation effect could attain as large as 0.3 for $\Delta 
P(\nu_\mu\to\nu_\tau) \equiv P(\nu_\mu\to\nu_\tau) - P(\bar{\nu_\mu}
\to\bar{\nu_\tau})$ and that the matter effect is negligibly small such as at most 
0.01 for $\Delta P(\nu_\mu\to\nu_\tau)$ for $\Delta m^2 = (1-5)\times 
10^{-3}$ ${\rm eV}^2$, which is the mass-squared difference relevant to the 
long-baseline oscillation.
\newpage

\centerline{\large\bf I  Introduction}
\vskip 0.2truecm

It has long been assumed that neutrinos are massless. However, since the 
atmospheric neutrino anomaly was discovered by several experimental 
Collaborations\cite{Kamiokande} and was affirmatively confirmed 
by the Super-Kamiokande\cite{SuperKamiokande}, people have come 
to think through the neutrino oscillation interpretation for the anomaly 
that neutrinos seem to have a certain amount of mass. Together with 
the solar neutrino deficit\cite{Solar}, the anomaly has been analyzed 
in the three-neutrino model\cite{Threenu} and two typical mass scales 
have been derived for the neutrino mass-squared difference; $\Delta 
m^2_{{\rm atm}} = (0.5 - 6)\times 10^{-3}$ ${\rm eV}^2$ with 
a large mixing angle of $\sin ^22\theta_{{\rm atm}}>0.82$ as the 
$\nu_\mu\to\nu_\tau$ oscillation from the atmospheric neutrino 
anomaly\cite{SuperKamiokande} and $\Delta m^2_{{\rm solar}} 
= (10^{-11} - 10^{-5})$ ${\rm eV}^2$, 
a large range depending on the three solutions of the 
vacuum oscillation and the MSW solutions in the matter with small- and 
large-angle mixings from the solar neutrino deficit\cite{Bahcall}. 

As in the quark sector, CP violation would be a characteristic feature in the 
three-neutrino model. It has been shown\cite{Tanimoto97}\cite{Arafune}
\cite{Minakata}\cite{Schubert} by using the constraints on the mixing 
matrix elements obtained from the analyses of these anomalies along with 
the results from the other 
accelerator and reactor experiments that the CP violation effect, 
defined as a difference of the oscillation probabilities between the neutrino 
and the antineutrino, is typically $1-3\%$ even in the long-baseline neutrino 
oscillations, depending on the assumed mass hierarchies.

On the other hand, sterile neutrinos were considered in the context of 
neutrino oscillations\cite{Barger80}\cite{Shi}. After that, a four-neutrino 
model of the ordinary three active neutrinos and one sterile neutrino was 
introduced with a mass pattern of two nearly degenerate pairs separated 
with a mass gap of the order of 1eV motivated from the hot dark matter
\cite{Peltoniemi} and then, by using the only one possible positive 
evidence from the terrestrial LSND experiments on the oscillations 
$\nu_\mu\to\nu_e$ and $\bar{\nu_\mu}\to\bar{\nu_e}$
\cite{LSND}, which suggest the mass scale of 
$\Delta m^2_{{\rm LSND}} = (0.3-2.2)$ ${\rm eV}^2$, 
the four-neutrino model with the same mass pattern as the above is studied
[14-20]. In this model, 
a sizable CP violation effect is shown to be possible in the long-baseline 
experiments\cite{Bilenky98}\cite{Dick}, and different magnitudes 
of the probability difference between the CP-conjugate channels are expected 
in between the three-neutrino model and the four-neutrino model
\cite{Tanimoto99} by using the most general parametrization of the mixing 
matrix\cite{Barger}. And, some features of CP asymmetry defined as 
the normalized probability difference are discussed in the long-baseline 
experiments at a neutrino factory\cite{Donini}. 

We will investigate the CP violation effect in the long-baseline neutrino 
oscillations numerically in more detail in the four-neutrino model with mass 
scheme of the two nearly degenerate pairs separated with the order of 
1 eV by using the most general parametrization 
of the mixing matrix, and in addition we will study the matter effect in the 
four-neutrino model. 

The paper is organized as follows. In Sec. II the four-neutrino model we use 
here is presented and the expressions of the difference of oscillation 
probabilities between the CP-conjugate channels are given both in the exact 
form and in the approximate forms relevant to the short-baseline and the 
long-baseline neutrino oscillations for the neutrino mass scheme mentioned above. 
In Sect. III constraints on the neutrino mixing matrix are derived by using 
the solar neutrino deficit, atmospheric neutrino anomaly, Bugey reactor 
experiment, CHOOZ experiment, LSND experiments, CHORUS 
and NOMAD experiments and the other accelerator and reactor experiments.
In Sect. IV the most general parametrization of the mixing matrix is adopted 
to obtain the constraints on the mixing angles and phases from the ones on 
the mixing matrix derived in Sect. III. And then, CP-violation in the 
long-baseline neutrino oscillation is investigated on the basis of these constraints. 
The behavior of the oscillation probability differences is analyzed in detail 
with respect to the two relevant phases of the mixing matrix and 
$\Delta m^2/E$. The matter effect 
is shown to be negligibly small in the four-neutrino model with the mass 
scheme adopted here. Finally, Sect. V is devoted to the conclusion.
\vskip 0.7truecm

\centerline{\large\bf II  The four-neutrino model}
\vskip 0.2truecm

In order to consider the solar neutrino deficit, the atmospheric neutrino 
anomaly and the LSND experiment, we will take the four-neutrino model with 
the three ordinary active neutrinos and one sterile neutrino with three different 
scales of the neutrino mass-squared difference, $\Delta m^2_{{\rm solar}} = 
(10^{-11} - 10^{-5})$ ${\rm eV}^2, \Delta m^2_{{\rm atm}} 
= (10^{-3} - 10^{-2})$ ${\rm eV}^2$ and $\Delta m^2_{{\rm LSND}} 
= (0.3 - 10)$ ${\rm eV}^2$.

Under the neutrino oscillation hypothesis\cite{Maki}\cite{Pontecorvo}, 
the flavor eigenstates of neutrinos are the mixtures of mass eigenstates with
masses $m_i  (i = 1, 2, 3, 4)$ as follows,
\begin{equation}
\nu_\alpha = \sum_{i=1}^4 U_{\alpha i}\nu_i, \qquad \alpha = e, \mu, \tau, s
\label{shiki1}
\end{equation}

\noindent
where $\nu_e, \nu_{\mu}$ and $\nu_\tau$ are the ordinary neutrinos and 
$\nu_s$ is the sterile neutrino, and $U$ is the unitary mixing 
matrix. The neutrino oscillation probability of $\nu_{\alpha} \to \nu_{\beta}$ 
in vacuum is given in the usual manner in the four-neutrino model by
\begin{equation}
P(\nu_{\alpha}\to\nu_{\beta}) = \delta_{\alpha\beta} - 4\sum_{k>j}
{\rm Re}(U^*_{\alpha k}U_{\alpha j}U^*_{\beta j}U_{\beta k})
\sin^2\Delta_{kj} + 2\sum_{k>j}{\rm Im}(U^*_{\alpha k}U_{\alpha j}
U^*_{\beta j}U_{\beta k})\sin 2\Delta_{kj},
\label{shiki2}
\end{equation}

\noindent
where $\Delta_{kj}\equiv\Delta m^2_{kj}L/(4E)$, $L$ being the distance from 
the neutrino source and $E$ the energy of neutrino. The oscillation probability 
for the antineutrinos is given by the exchange of $U\leftrightarrow U^*$ in 
Eq.(\ref{shiki2}). And, the probability difference between CP-conjugate 
channels given by 
\begin{eqnarray}
\Delta P_{\alpha\beta} &\equiv& P(\nu_{\alpha}\to\nu_{\beta}) - 
P(\bar{\nu_{\alpha}}\to\bar{\nu_{\beta}})  \nonumber  \\
&=& 4\sum_{k>j} {\rm Im}(U^*_{\alpha k}U_{\alpha j}
U^*_{\beta j}U_{\beta k})\sin 2\Delta_{kj}    \label{shiki3}
\end{eqnarray}
is a direct measure of the genuine CP-violation effect in the neutrino 
oscillation in vacuum\cite{Cabibbo}. 

The four neutrino masses should be devided into two pairs of close masses 
separated by a gap of about 1eV in order to accomodate with 
the solar and atmospheric neutrino deficits and the LSND experiments 
along with the other results from the accelerator and reactor experiments 
on the neutrino oscillation. There are the following two schemes for that 
mass pattern; (i) $\Delta m^2_{{\rm solar}} \equiv \Delta m^2_{21} \ll 
\Delta m^2_{{\rm atm}} \equiv \Delta m^2_{43} \ll 
\Delta m^2_{{\rm LSND}} \equiv \Delta m^2_{32}$, and (ii) 
$\Delta m^2_{{\rm solar}} \equiv \Delta m^2_{43} \ll 
\Delta m^2_{{\rm atm}} \equiv \Delta m^2_{21} \ll 
\Delta m^2_{{\rm LSND}} \equiv \Delta m^2_{32}$, 
where $\Delta m^2_{kj} \equiv m^2_k - m^2_j$. We will 
adopt the first scheme in the following analyses, and the second scheme can 
be attained only through the exchange of indices $(1, 2)\leftrightarrow (3, 4)$ 
in the following various expressions such as the oscillation probabilities. 
In the first scheme, the measure of CP violation in the neutrino oscillation 
in vacuum is given for the short-baseline experiment 
( $L/E \sim 1$ [km/GeV] ) as follows, 
\begin{eqnarray}
\Delta P_{\alpha\beta} &\simeq& 4[ {\rm Im}(U^*_{\alpha 3}U_{\alpha 2}
U^*_{\beta 2}U_{\beta 3}) + {\rm Im}(U^*_{\alpha 3}U_{\alpha 1}
U^*_{\beta 1}U_{\beta 3})  \nonumber  \\
& & {} +{\rm Im}(U^*_{\alpha 4}U_{\alpha 2}U^*_{\beta 2}U_{\beta 4}) 
+ {\rm Im}(U^*_{\alpha 4}U_{\alpha 1}U^*_{\beta 1}U_{\beta 4}) ] 
\sin 2\Delta_{32} ,   \label{shiki4}
\end{eqnarray}
since $\Delta_{21}$ and $\Delta_{43} \ll 1$, and $\Delta_{41}, \Delta_{42}, 
\Delta_{31}, \Delta_{32} \simeq 1$. $\Delta P_{\alpha\beta}$ in 
Eq.(\ref{shiki4}) is zero due to the unitarity of the mixing matrix $U$. So, 
CP violation is negligibly small in the short-baseline oscillation experiments 
in the four-neutrino model. 

On the other hand, for the long-baseline experiment ( $L/E = 100 - 1000$ 
[km/GeV]) the probability difference in vacuum is given as follows, 
\begin{equation}
\Delta P_{\alpha\beta} \simeq 4{\rm Im}(U^*_{\alpha 4}U_{\alpha 3}
U^*_{\beta 3}U_{\beta 4})\sin 2\Delta_{43} ,  \label{shiki5}
\end{equation}
since $\Delta_{21} \ll 1$, $\Delta_{41}, \Delta_{42}, \Delta_{31}, 
\Delta_{32} \gg 1$, and $\Delta_{43} \sim 1$. There are six 
$\Delta P_{\alpha\beta}$'s, that is, $\Delta P_{\mu e}, \Delta P_{e\tau}, 
\Delta P_{\mu\tau}, \Delta P_{es}, \Delta P_{\mu s}$, and 
$\Delta P_{\tau s}$. Three of these six $\Delta P_{\alpha\beta}$'s are 
independent due to the unitarity of $U$ for the approximate expression 
of $\Delta P_{\alpha\beta}$ in Eq.(\ref{shiki5}) as well as for the exact 
expression in Eq.(\ref{shiki3}). 
\vskip 0.7truecm

\newpage
\centerline{\large\bf III  Constraints on the mixing matrix $U$}
\vskip 0.2truecm

In order to numerically calculate the oscillation probability differences 
$\Delta P_{\alpha\beta}$, we will derive the constraints on the mixing 
matrix $U$ from the solar neutrino deficit, atmospheric neutrino 
anomaly, LSND experiments and the other terrestrial oscillation 
experiments using the accelerators and reactors. 

\noindent
(i) Solar neutrino deficit

\noindent
Since $\Delta_{21} \sim 1$ and all the other five $\Delta_{kj}$'s are 
enormously larger than 1, the survival probability of $\nu_e$ is given 
from Eq.(\ref{shiki2}) by 

\begin{eqnarray}
P_{{\rm solar}}(\nu_e\to\nu_e) &\simeq& 1-4|U_{e1}|^2|U_{e2}|^2
\sin^2\Delta_{21} - 2|U_{e3}|^2(1-|U_{e3}|^2-|U_{e4}|^2)  \nonumber  \\
& & {} -2|U_{e4}|^2(1-|U_{e4}|^2) ,  \label{shiki6}
\end{eqnarray}
where the unitarity of $U$ is used. For the solar neutrino deficit, there are 
three different kinds of solutions, that is, the vacuum solution and the MSW 
solutions with small and large angle mixings, and a unique solution is not yet 
found, so that we will not use this deficit in order to obtain the constraints. 

\noindent
(ii) Atmospheric neutrino anomaly

\noindent
Since $\Delta_{21} \ll 1, \Delta_{43} \sim 1$ and $\Delta_{41}, \Delta_{42}, 
\Delta_{31}, \Delta_{32} \gg 1$, the survival probability of $\nu_{\mu}$ 
is given by 
\begin{equation}
P_{{\rm atm}}(\nu_{\mu}\to\nu_{\mu}) \simeq 1 - 4|U_{\mu 3}|^2
|U_{\mu 4}|^2\sin^2\Delta_{43} - 2(|U_{\mu 1}|^2+|U_{\mu 2}|^2)
(1-|U_{\mu 1}|^2-|U_{\mu 2}|^2).  \label{shiki7} 
\end{equation}
By using the data from the Super-Kamiokande experiments, that is, 
$\sin^22\theta_{{\rm atm}}>0.82$ for $5\times 10^{-4}<
\Delta m^2_{{\rm atm}}<6\times 10^{-3}$ ${\rm eV}^2$, and expecting 
from this data that $|U_{\mu 1}|^2+|U_{\mu 2}|^2\ll 1$, the following 
constraint is obtained, 
\begin{equation}
|U_{\mu 3}|^2|U_{\mu 4}|^2 > 0.205 .  \label{shiki8}
\end{equation}

\noindent
(iii) The Bugey experiment\cite{Bugey} (including Krasnoyarsk\cite{Krasno}, 
CDHS\cite{CDHS} and CCFR\cite{CCFR} experiments)

\noindent
By being typically represented by the Bugey reactor experiment with $L/E=3-20$ 
[m/MeV or km/GeV], since $\Delta_{21} \ll 1, \Delta_{43} \ll 1$ and 
$\Delta_{41}, \Delta_{42}, \Delta_{31}, \Delta_{32} \sim 1$, the survival 
probability of $\bar{\nu_e}$ is given by 
\begin{equation}
P_{{\rm Bugey}}(\bar{\nu_e}\to\bar{\nu_e}) \simeq 1 - 4(|U_{e3}|^2
+|U_{e4}|^2)(1-|U_{e3}|^2-|U_{e4}|^2)\sin^2\Delta_{32}.  \label{shiki9} 
\end{equation}
If we use the data from the Bugey experiment conservatively, that is, 
$\sin^22\theta_{{\rm Bugey}}<0.1$ for $0.1<\Delta m^2<
1$ ${\rm eV}^2$, the following constraint is obtained, 
\begin{equation}
|U_{e3}|^2+|U_{e4}|^2 < 0.025 .  \label{shiki10}
\end{equation}

\noindent
(iv) The CHOOZ experiment\cite{CHOOZ}

\noindent
This experiment is the first long-baseline reactor experiment, since 
$L\sim 1$ km and $E\sim 3$ MeV so that $L/E\sim 300$ [km/GeV]. 
Therefore, $\Delta_{21} \ll 1, \Delta_{43} \sim 1$ and $\Delta_{41}, 
\Delta_{42}, \Delta_{31}, \Delta_{32} \gg 1$, and the survival probability of 
$\bar{\nu_e}$ is given by 
\begin{equation}
P_{{\rm CHOOZ}}(\bar{\nu_e}\to\bar{\nu_e}) \simeq 1 - 4|U_{e3}|^2
|U_{e4}|^2\sin^2\Delta_{43} - 2(|U_{e3}|^2+|U_{e4}|^2)
(1-|U_{e3}|^2-|U_{e4}|^2).  
\label{shiki11} 
\end{equation}
By using the data from the CHOOZ experiment, that is, 
$\sin^22\theta_{{\rm CHOOZ}}<0.12$ for $3\times 10^{-3}
<\Delta m^2<1.0\times 10^{-2}$ ${\rm eV}^2$ and adopting 
Eq.(\ref{shiki10}), the following constraint is obtained, 
\begin{equation}
4|U_{e3}|^2|U_{e4}|^2 < 0.12 .  \label{shiki12}
\end{equation}
If we use, however, the constraint of Eq.(\ref{shiki10}) and the unequality of 
$2|U_{e3}||U_{e4}|\leq |U_{e3}|^2+|U_{e4}|^2$, a constraint 
$4|U_{e3}|^2|U_{e4}|^2 <6.3\times 10^{-4}$ is obtained so that 
Eq.(\ref{shiki12}) is included in the constraint from the Bugey experiment. 

\noindent
(v) The LSND experiments\cite{LSND}

\noindent
This experiment is of the short baseline, $L/E=0.5-1$ [m/MeV]. 
Since $\Delta_{21} \ll 1, \Delta_{43} \ll 1$ and 
$\Delta_{41}, \Delta_{42}, \Delta_{31}, \Delta_{32} \sim 1$, 
the oscillation probability of $\nu_{\mu}\to\nu_e$ is expressed as follows, 
\begin{eqnarray}
P_{{\rm LSND}}(\nu_{\mu}\to\nu_e) &\simeq& -4{\rm Re}\left[ 
(U^*_{\mu 3}U_{e3}+U^*_{\mu 4}U_{e4})(U_{\mu 1}U^*_{e1}
+U_{\mu 2}U^*_{e2})\right]\sin^2\Delta_{32}  \nonumber  \\
&=& 4|U^*_{\mu 3}U_{e3}+U^*_{\mu 4}U_{e4}|^2\sin^2\Delta_{32} , 
\label{shiki13}
\end{eqnarray}
where the unitarity of $U$ is used. By using the data from the LSND 
experiments, that is, $\sin^22\theta_{{\rm LSND}}=1.5\times 10^{-3}
-1.0\times 10^{-1}$ for $0.3<\Delta m^2_{{\rm LSND}}<2.2$ 
${\rm eV}^2$, the following constraint is obtained, 
\begin{equation}
|U^*_{\mu 3}U_{e3} +U^*_{\mu 4}U_{e4}| = 0.02 - 0.16 .  \label{shiki14}
\end{equation}

\noindent
(vi) The CHORUS\cite{CHORUS} and NOMAD\cite{NOMAD} experiments

\noindent
These experiments are also the short baseline ones searching for the
 $\nu_{\mu}\to\nu_{\tau}$ oscillation, $L/E=0.02-0.03$ [km/GeV]. 
Since $\Delta_{21} \ll 1, \Delta_{43} \ll 1$ and $\Delta_{41}, \Delta_{42}, 
\Delta_{31}, \Delta_{32} \simeq 10^{-2}-10^{-1}$, the oscillation 
probability is given by 
\begin{equation}
P_{{\rm CHORUS/NOMAD}}(\nu_{\mu}\to\nu_{\tau}) \simeq 
4|U^*_{\mu 3}U_{\tau 3} + U^*_{\mu 4}U_{\tau 4}|^2\sin^2\Delta_{32} . 
\label{shiki15}
\end{equation}
By using the data from the latest NOMAD experiment, 
$\sin^22\theta_{{\rm NOMAD}}<0.3$ for $\Delta m^2<2.2$ ${\rm eV}^2$, 
the following constraint is obtained, 
\begin{equation}
|U^*_{\mu 3}U_{\tau 3} +U^*_{\mu 4}U_{\tau 4}| < 0.28 .  \label{shiki16}
\end{equation}

Among the above-mentioned six typical phenomena and experiments, 
the useful constraints are of Eqs. (\ref{shiki8}), (\ref{shiki10}), 
(\ref{shiki14}) and (\ref{shiki16}). 
\vskip 0.7truecm

\centerline{\large\bf IV. CP violation in the neutrino oscillations}
\vskip 0.2truecm

In this section, by using the constraints obtained in the previous section, 
we will numerically investigate the CP violation effects in the long-baseline 
neutrino oscillation experiments in the four-neutrino model described in Sect. II. 

We adopt the most general parametrization of the mixing matrix $U$ 
for Majorana neutrinos\cite{Barger}, which includes six mixing angles and 
six phases. The expression of the matrix is too complicated to write it down here. 
So, we cite only the matrix elements which are useful for the following 
numerical analyses; $U_{e1}=c_{01}c_{02}c_{03}, U_{e2}=
c_{02}c_{03}s^*_{d01}, U_{e3}=c_{03}s^*_{d02}, U_{e4}=s^*_{d03}, 
U_{\mu 3}=-s^*_{d02}s_{d03}s^*_{d13}+c_{02}c_{13}s^*_{d12}, 
U_{\mu 4}=c_{03}s^*_{d13}, U_{\tau 3}=-c_{13}s^*_{d02}s_{d03}
s^*_{d23}-c_{02}s^*_{d12}s_{d13}s^*_{d23}+c_{02}c_{12}c_{23}$, and 
$U_{\tau 4}=c_{03}c_{13}s^*_{d23}$, where $c_{ij}\equiv \cos\theta_{ij}$ 
and $s_{dij}\equiv s_{ij}{\rm e}^{{\rm i}\delta_{ij}}\equiv \sin\theta_{ij}
{\rm e}^{{\rm i}\delta_{ij}}$ \cite{Barger}, and $\theta_{01}, \theta_{02}, 
\theta_{03}, \theta_{12}, \theta_{13}, \theta_{23}$ are the six angles and 
$\delta_{01}, \delta_{02}, \delta_{03}, \delta_{12}, \delta_{13}, 
\delta_{23}$ are the six phases. As stated in Sect. II, three of the six oscillation 
probability differences are independent so that only three of the six phases are 
determined by the measurements of the CP violation effect in the neutrino 
oscillations. In this sense, our analyses apply both to the Dirac and Majorana 
neutrinos\cite{Tanimoto99}. 

On the basis of this parametrization, we obtain the constraints on the mixing 
angles and phases by using the constraints on the mixing matrix elements 
derived in the previous section. First, the constraint of Eq.(\ref{shiki10}) 
leads to 
\begin{equation}
c^2_{03}s^2_{02} + s^2_{03} < 0.025 .  \label{shiki17}
\end{equation}
This unequality means at least $s^2_{02}, s^2_{03}<0.025$. The next 
constraint of Eq.(\ref{shiki8}) of 
\begin{equation}
 | -s_{02}s_{03}s_{13}{\rm e}^{-{\rm i}(\delta_{02}-\delta_{03}
+\delta_{13})}+c_{02}c_{13}s_{12}{\rm e}^{-{\rm i}\delta_{12}} |^2
c^2_{03}s^2_{13} > 0.205   
\end{equation}
leads to 
\begin{equation}
s^2_{12}c^2_{13}s^2_{13} > 0.205   \label{shiki18}
\end{equation}
due to the smallness of $s_{02}$ and $s_{03}$. The third constraint of 
Eq.(\ref{shiki14}) gives the following expression, 
\begin{equation}
 | c_{02}s_{02}c_{03}s_{12}c_{13} + c^2_{02}c_{03}s_{03}s_{13}
{\rm e}^{{\rm i}\delta_1}| = 0.02 - 0.16 ,  \label{shiki19}
\end{equation}
where $\delta_1\equiv \delta_{02}-\delta_{03}-\delta_{12}+\delta_{13}$. 
This constraint proves not to bring any constraint on the phase $\delta_1$, 
if we use Eqs.(\ref{shiki17}) and (\ref{shiki18}). The fourth constraint of 
Eq.(\ref{shiki16}) is expressed as 
\begin{eqnarray}
&|&c^2_{02}c_{12}s_{12}c_{13}c_{23}-c_{02}s_{02}s_{03}s_{12}
c^2_{13}s_{23}{\rm e}^{-{\rm i}(\delta_1+\delta_2)}-c_{02}s_{02}
s_{03}c_{12}s_{13}c_{23}{\rm e}^{{\rm i}\delta_1}   \nonumber  \\
& & {} +c_{13}s_{13}s_{23}(c^2_{03}-c^2_{02}s^2_{12}
+s^2_{02}s^2_{03}){\rm e}^{-{\rm i}\delta_2}| < 0.28 ,   \label{shiki20}
\end{eqnarray}
where $\delta_2\equiv \delta_{12}-\delta_{13}+\delta_{23}$. By using 
Eqs. (\ref{shiki17}) and (\ref{shiki18}), no constraint on $\delta_1$, 
and the fact of the large angle mixing in $\nu_{\mu}\to\nu_{\tau}$ oscillation 
for the atmospheric neutrino anomaly which leads to the nearly maximal mixing 
in the angle $\theta_{23}$, the constraint of Eq. (\ref{shiki20}) gives no 
constraint to the phase $\delta_2$. 

So, in summary, we derive the two constraints of Eqs. (\ref{shiki17}) and 
(\ref{shiki18}) on the mixing angles and no constraints on the two phases of 
$\delta_1$ and $\delta_2$. 

Using these two constraints on the mixing angles, 
we will calculate the differences of the oscillation probabilities between the 
CP-conjugate channels for the long-baseline neutrino oscillations. As stated 
before, only three of the six probability differences among the four neutrinos 
are independent so that three of the six phases are relevant here. 
However, only two phases dominantly affect the differences as is shown 
by the leading terms relevant to the long-baseline oscillation, which are given 
in the following, 
\begin{eqnarray}
\Delta P_{\mu e}&\simeq&4c^2_{03}c_{02}s_{02}s_{03}s_{12}
c_{13}s_{13}\sin\delta_1\sin 2\Delta_{43} ,   \nonumber  \\
\Delta P_{e\tau}&\simeq& 4c^2_{03}c_{02}s_{02}s_{03}c_{13}s_{23}
\left[ -c_{12}c_{23}\sin (\delta_1+\delta_2)+s_{12}s_{13}s_{23}
\sin\delta_1\right] \sin 2\Delta_{43} ,  \nonumber  \\
\Delta P_{\mu\tau}&\simeq& 4c^2_{03}c_{02}c_{13}s_{13}s_{23}
[ c_{02}c_{12}s_{12}c_{13}c_{23}\sin\delta_2+s_{02}s_{03}s_{12}
s_{23}\sin\delta_1   \nonumber  \\
& & {} -s_{02}s_{03}c_{12}s_{13}c_{23}\sin (\delta_1+\delta_2) ] 
\sin 2\Delta_{43} ,   \label{shiki21}  \\
\Delta P_{es}&\simeq& 4c^2_{03}c_{02}s_{02}s_{03}c_{13}c_{23}
\left[ c_{12}s_{23}\sin (\delta_1+\delta_2)+s_{12}s_{13}c_{23}
\sin\delta_1\right] \sin 2\Delta_{43} ,  \nonumber  \\
\Delta P_{\mu s}&\simeq& 4c^2_{03}c_{02}c_{13}s_{13}c_{23}
[ c_{02}c_{12}s_{12}c_{13}s_{23}\sin\delta_2-s_{02}s_{03}s_{12}
c_{23}\sin\delta_1   \nonumber  \\
& & {} -s_{02}s_{03}c_{12}s_{13}s_{23}\sin (\delta_1+\delta_2) ] 
\sin 2\Delta_{43} ,  \nonumber  \\
\Delta P_{\tau s}&\simeq& -4c^2_{03}c_{02}c_{12}c^2_{13}c_{23}s_{23}
\left[ c_{02}s_{12}s_{13}\sin\delta_2+s_{02}s_{03}c_{13}\sin (\delta_1
+\delta_2)\right] \sin 2\Delta_{43} ,   \nonumber 
\end{eqnarray}
where $\delta_1$ and $\delta_2$ are the linear combinations of $\delta_{ij}$'s 
as stated before. We take the range of phases as $0\leq \delta_{ij}<2\pi$ and 
the range of mixing angles as $0\leq \theta_{ij}\leq\pi$ so that $s_{ij}$'s can be 
taken only positive and $c_{ij}$'s can be taken both positive and negative. 
Since Eq. (\ref{shiki17}) means that the angles $\theta_{02}$ and 
$\theta_{03}$ are very small and Eq. (\ref{shiki18}) leads to
 $s^2_{12}\sin^22\theta_{13}>0.82$ which means that $s_{12}$ is 
in the range of $0.9\leq s_{12}\leq 1.0$ and the angle $\theta_{13}$ is 
around $\pi /4$, $\Delta P_{\mu e}$ and $\Delta P_{e\tau}$ are expected 
from Eq.(\ref{shiki21}) to be very small and $\Delta P_{\mu\tau}$ is to be 
able to take a sizable magnitude. In the following, we calculate the 
oscillation probabilities $P(\nu_{\alpha}\to\nu_{\beta})$ and their 
differences $\Delta P_{\alpha\beta}$ by using the rigorous expressions of 
Eqs. (\ref{shiki2}) and (\ref{shiki3}). 

\begin{table}
\caption{Phase $\delta_1$-dependence of the oscillation probabilities 
$P_{\alpha\beta}$ and their differences $\Delta P_{\alpha\beta}$ for the 
long-baseline experiment. $P_{\alpha\beta}\equiv P(\nu_{\alpha}\to
\nu_{\beta})$ and $\Delta P_{\alpha\beta}\equiv  P(\nu_{\alpha}\to
\nu_{\beta})-P(\bar{\nu_{\alpha}}\to\bar{\nu_{\beta}})$. $\delta_1$ is 
in degree.}
\begin{center}
\setlength{\tabcolsep}{12pt}
\begin{tabular}{rrrrrrr} \hline
\multicolumn{1}{c}{$\delta_1$} & 
\multicolumn{1}{c}{$P_{\mu e}$} & 
\multicolumn{1}{c}{$P_{e\tau}$} & 
\multicolumn{1}{c}{$P_{\mu\tau}$} & 
\multicolumn{1}{c}{$\Delta P_{\mu e}$} & 
\multicolumn{1}{c}{$\Delta P_{e\tau}$} & 
\multicolumn{1}{c}{$\Delta P_{\mu\tau}$}   \\ \hline 
$0^\circ$ & 0.032 & 0.004 & 0.148 & 0.000 & -0.007 & -0.269 \\
$45^\circ$ & 0.036 & 0.008 & 0.148 & 0.015 & 0.002 & -0.277 \\
$90^\circ$ & 0.032 & 0.015 & 0.147 & 0.022 & 0.011 & -0.283 \\
$135^\circ$ & 0.022 & 0.021 & 0.147 & 0.015 & 0.013 & -0.282 \\
$180^\circ$ & 0.011 & 0.022 & 0.148 & 0.000 & 0.007 & -0.275 \\
$225^\circ$ & 0.006 & 0.018 & 0.148 & -0.015 & -0.002 & -0.267 \\
$270^\circ$ & 0.011 & 0.011 & 0.148 & -0.022 & -0.011 & -0.261 \\
$315^\circ$ & 0.021 & 0.005 & 0.148 & -0.015 & -0.013 & -0.262 \\ \hline
\end{tabular}
\end{center}
\label{tab1}
\end{table}
\vskip 0.1truecm

\begin{table}
\caption{Phase $\delta_2$-dependence of the oscillation probabilities 
$P_{\alpha\beta}$ and their differences $\Delta P_{\alpha\beta}$ for the 
long-baseline experiment. $\delta_2$ is in degree.}
\begin{center}
\setlength{\tabcolsep}{12pt}
\begin{tabular}{rrrrrrr} \hline
\multicolumn{1}{c}{$\delta_2$} & 
\multicolumn{1}{c}{$P_{\mu e}$} & 
\multicolumn{1}{c}{$P_{e\tau}$} & 
\multicolumn{1}{c}{$P_{\mu\tau}$} & 
\multicolumn{1}{c}{$\Delta P_{\mu e}$} & 
\multicolumn{1}{c}{$\Delta P_{e\tau}$} & 
\multicolumn{1}{c}{$\Delta P_{\mu\tau}$}   \\ \hline 
$0^\circ$ & 0.032 & 0.009 & 0.191 & 0.022 & 0.003 & -0.008 \\
$45^\circ$ & 0.032 & 0.009 & 0.124 & 0.022 & 0.006 & -0.201 \\
$90^\circ$ & 0.032 & 0.015 & 0.147 & 0.022 & 0.011 & -0.283 \\
$135^\circ$ & 0.032 & 0.023 & 0.248 & 0.022 & 0.016 & -0.205 \\
$180^\circ$ & 0.032 & 0.028 & 0.367 & 0.022 & 0.018 & -0.014 \\
$225^\circ$ & 0.032 & 0.028 & 0.433 & 0.022 & 0.016 & 0.179 \\
$270^\circ$ & 0.032 & 0.022 & 0.409 & 0.022 & 0.011 & 0.261 \\
$315^\circ$ & 0.032 & 0.014 & 0.309 & 0.022 & 0.006 & 0.184 \\ \hline
\end{tabular}
\end{center}
\label{tab2}
\end{table}
\vskip 0.1truecm

The probabilities of $P(\nu_{\mu}\to\nu_e)$ and $P(\bar{\nu_{\mu}}
\to\bar{\nu_e})$ as functions of the phase $\delta_1$ with $\delta_2=
\pi /2$ fixed are shown in Fig.1 and those of $P(\nu_e\to\nu_{\tau})$ 
and $P(\bar{\nu_e}\to\bar{\nu_{\tau}})$ as functions of the phase 
$\delta_2$ with $\delta_1=\pi /2$ fixed are shown in Fig.2 for the values 
of the parameter set of angles and phases; $s_{02}=s_{03}=0.11 
(c_{02}=c_{03}=0.994), s_{12}=0.91 (c_{12}=0.415), s_{13}=0.67 
(c_{13}=0.742), s_{01}=s_{23}=1/\sqrt{2} (c_{01}=c_{23}=1/\sqrt{2})$ 
and $\delta_{01}=\delta_{02}=\delta_{03}=\delta_{12}=0$, which are 
chosen so as to give the probability differences as large as possible within 
the parameter ranges allowed by the constaints of Eqs. (\ref{shiki17}) and 
(\ref{shiki18}). The magnitude of these probabilities is at most 0.04 
as shown in Figs.1 and 2. Therefore, the probability differences 
$\Delta P(\nu_{\mu}\to\nu_e)$ and $\Delta P(\nu_e\to
\nu_{\tau})$ are at most $\pm 0.02$ as shown in Fig.3 for the same 
parameter values. On the other hand, $P(\nu_{\mu}\to\nu_{\tau})$ 
and $P(\bar{\nu_{\mu}}\to\bar{\nu_{\tau}})$ can rise to as large as 
$0.40 - 0.45$ as shown in Fig.4 and $\Delta P(\nu_{\mu}\to\nu_{\tau})$ 
can attain as large as $\pm 0.28$ as shown in Fig.5 for the same parameter 
values. These facts agree with the above-mentioned expectations. 
The angle $\theta_{23}$-dependence of $\Delta P(\nu_{\mu}\to\nu_{\tau})$ 
and $\Delta P(\nu_e\to\nu_{\tau})$ is shown in Fig.6, where the phases 
$\delta_1$ and $\delta_2$ are taken as $\pi /2$ and the values of the other 
angles and phases are the same as the above. We display the phase 
$\delta_1$-dependence in Table 1 and the phase $\delta_2$-dependence 
in Table 2 of $P(\nu_{\mu}\to\nu_e), P(\nu_e\to\nu_{\tau}), 
P(\nu_{\mu}\to\nu_{\tau}), \Delta P(\nu_{\mu}\to\nu_e), 
\Delta P(\nu_e\to\nu_{\tau})$ and $\Delta P(\nu_{\mu}\to\nu_{\tau})$. 

Here we comment on the matter effect on the oscillation probability difference 
in the four-neutrino model. By using the 
Minakata-Nunokawa procedure\cite{Minakata}, the probability difference 
with the matter effect is expressed for the long-baseline 
$\nu_{\alpha}\to\nu_{\beta}$ oscillation in the four-neutrino model with 
mass scheme of the two nearly degenerate pairs separated with the order of 1 eV 
as follows, 
\begin{eqnarray}
\Delta P_{\alpha\beta}&\simeq&4{\rm Im}(U^*_{\alpha 4}U_{\alpha 3}
U^*_{\beta 3}U_{\beta 4})\cos B_{34}\sin\left( \frac{\Delta m^2L}{2E}
\right)  \nonumber  \\
& & {} +4{\rm Re}(U^*_{\alpha 4}U_{\alpha 3}U^*_{\beta 3}U_{\beta 4})
\sin B_{34}\sin\left( \frac{\Delta m^2L}{2E}\right)  \nonumber  \\
& & {} -8\sum_{j>i}{\rm Re}(UUU\delta V)_{\alpha\beta;ij}\cos^2\left( 
\frac{B_{ij}}{2}\right) \sin^2\left( \frac{\Delta m^2_{ij}L}{4E}\right) , 
\label{shiki22}
\end{eqnarray}
where 
\begin{equation}
B_{ij} = (|U_{ei}|^2-|U_{ej}|^2)aL + (|U_{si}|^2-|U_{sj}|^2)a'L , 
\label{shiki23}
\end{equation}
\begin{eqnarray}
(UUU\delta V)_{\alpha\beta;ij}&=&U^*_{\alpha i}U_{\alpha j}
U^*_{\beta j}\delta V_{\beta i}+U^*_{\alpha i}U_{\alpha j}
\delta V^*_{\beta j}U_{\beta i}  \nonumber  \\
& & {} +U^*_{\alpha i}\delta V_{\alpha j}U^*_{\beta j}U_{\beta i}
+\delta V^*_{\alpha i}U_{\alpha j}U^*_{\beta j}U_{\beta i} .
\label{shiki24}
\end{eqnarray}
In Eq.(\ref{shiki23}), the quantity $a$ represents the matter effect for 
$\nu_e$ and we take $a=1.04\times 10^{-13}$ eV for the constant matter 
density of 2.7 ${\rm g/cm}^3$ \cite{Minakata}, and $a'$ represents the one 
for $\nu_s$ and we take $a' = a/2$ \cite{Bilenky98}. In Eq.(\ref{shiki24}), 
$\delta V_{\alpha i}$ is given \cite{Minakata} by 
\begin{equation}
\delta V_{\alpha i} = \sum_{j\neq i}\frac{2E}{\Delta m^2_{ij}}
U_{\alpha j}(U^*_{ej}U_{ei}a + U^*_{sj}U_{si}a') .  \label{shiki25}
\end{equation}
In Eq.(\ref{shiki22}), the first term represents the genuine CP-violation effect 
corrected by the matter effect, the second term does the CP-violation effect coming 
from the phase evolution of the neutrino wave function in the matter, and the 
third one results from the corrections to the mixing matrix $U$ due to the existence 
of matter\cite{Minakata}. 

We estimate these matter effects for the $\nu_{\mu}\to\nu_{\tau}$ 
oscillation. The first term of Eq.(\ref{shiki22}) is almost the genuine 
CP-violation effect, since 
the magnitude of the matter effect $B_{34}$ is at most $1\times 10^{-3}$ 
for the above-mentioned parameter values of the mixing angles and phases. 
The second term is approximately $0.4\times 10^{-3}$, since 
${\rm Re}(U^*_{\mu 4}U_{\mu 3}U^*_{\tau 3}U_{\tau 4})\sim 0.1$ 
and $\sin B_{34}\sim 1\times 10^{-3}$. This should be compared with 
the possible maximum value of the genuine CP-violation effect displayed 
in Fig.5 and Table 1, that is, $|\Delta P(\nu_{\mu}\to\nu_{\tau})| \sim 0.3$. 
The third term of Eq.(\ref{shiki22}) is expressed as 
\begin{equation}
-8\sum_{i=1,2,j=3,4}{\rm Re}(UUU\delta V)_{\mu\tau;ij}\sin^2\left( 
\frac{\Delta M^2L}{4E}\right) - 8{\rm Re}(UUU\delta V)_{\mu\tau;34}
\sin^2\left( \frac{\Delta m^2L}{4E}\right) ,  \label{shiki26}
\end{equation}
since $\cos^2\left( B_{ij}/2\right) \simeq 1.0$. The coefficient 
of the first term of Eq.(\ref{shiki26}) is given by 
\begin{eqnarray}
\sum_{i=1,2,j=3,4}&{\rm Re}&(UUU\delta V)_{\mu\tau;ij}  \nonumber  \\
&=&4\frac{2Ea}{\Delta M^2}{\rm Re}[ (U^*_{\mu 3}U_{\tau 3}
+U^*_{\mu 4}U_{\tau 4})\{ (U_{\mu 3}U^*_{e3}+U_{\mu 4}U^*_{e4})
(U^*_{\tau 3}U_{e3}+U^*_{\tau 4}U_{e4})  \nonumber  \\
& & {} +\frac{1}{2}(U_{\mu 3}U^*_{s3}+U_{\mu 4}U^*_{s4})
(U^*_{\tau 3}U_{s3}+U^*_{\tau 4}U_{s4}) \} ] ,  \label{shiki27} 
\end{eqnarray}
where the relation $a'=a/2$ is used, and the terms with $1/\Delta m^2$ 
and $1/\Delta m^2_{{\rm solar}}$ do not 
appear due to the symmetry of the mass scheme of the four neutrinos adopted 
in our model. The coefficient of the second term of Eq.(\ref{shiki26}) 
is given by 
\begin{eqnarray}
&{\rm Re}&(UUU\delta V)_{\mu\tau;34}  \nonumber  \\
& & {} =-\frac{2Ea}{\Delta M^2}
{\rm Re}[ \{ U_{\mu 3}U^*_{\mu 4}(|U_{\tau 3}|^2+|U_{\tau 4}|^2)
+U_{\tau 3}U^*_{\tau 4}(|U_{\mu 3}|^2+|U_{\mu 4}|^2) \} 
(U^*_{e3}U_{e4}  \nonumber  \\
& & {} +\frac{1}{2}U^*_{s3}U_{s4})  +2U^*_{\mu 3}U_{\mu 4}
U_{\tau 3}U^*_{\tau 4}(|U_{e3}|^2+|U_{e4}|^2 +\frac{1}{2}|U_{s3}|^2 
+\frac{1}{2}|U_{s4}|^2) ]   \nonumber  \\
& & {} +\frac{2Ea}{\Delta m^2}{\rm Re}[ \{ U_{\mu 3}U^*_{\mu 4}
(|U_{\tau 3}|^2-|U_{\tau 4}|^2)+U_{\tau 3}U^*_{\tau 4}
(|U_{\mu 3}|^2-|U_{\mu 4}|^2) \}(U^*_{e3}U_{e4}  \nonumber  \\
& & {} +\frac{1}{2}U^*_{s3}U_{s4}) ] .  \label{shiki28}
\end{eqnarray}
The magnitude of Eq.(\ref{shiki27}) is estimated to be $-0.9\times 10^{-4}$, 
and the magnitude of Eq.(\ref{shiki28}) is to be $1.1\times 10^{-2}, 
0.51\times 10^{-2}, 1.3\times 10^{-3}$ for $\Delta m^2=(1.0, 2.0, 5.0)
\times 10^{-3}$ ${\rm eV}^2$, respectively. So, the third term of 
Eq.(\ref{shiki22}), that is, Eq.(\ref{shiki26}) for $\nu_\mu\to\nu_\tau$ 
oscillation is again negligibly 
small as compared with the possible maximum value of the genuine 
CP-violation effect. So, the matter effect can be totally neglected in the 
$\nu_\mu\to\nu_\tau$ oscillation in the four-neutrino model with mass 
scheme of the two nearly degenerate pairs separated with the order of 
$1{\rm eV}$, as was generally studied for any channels 
in ref.\cite{Tanimoto99}. 

As can be seen in Figs.5 and 6 and in Tables 1 and 2, CP violation could be 
observed as the probability difference between the $\nu_{\mu}\to\nu_\tau$ 
and $\bar{\nu_\mu}\to\bar{\nu_\tau}$ oscillations in the four-neutrino 
model. So, we show in Figs.7 and 8 the oscillation probabilities 
$P(\nu_\mu\to\nu_\tau)$ and $P(\bar{\nu_\mu}\to\bar{\nu_\tau})$, 
and their difference $\Delta P(\nu_\mu\to\nu_\tau)$ as functions of 
$\Delta m^2/E$ $[{\rm eV}^2/{\rm GeV}]$, respectively, for the 
long-baseline experiments of the MINOS\cite{MINOS} and CERN-ICARUS
\cite{ICARUS} types, where $E$ is the neutrino energy. In Figs.7 and 8, 
we have assumed the baseline length as $L = 730$ km. We can observe 
from Fig.8 that if the beam energy is taken as 7 GeV, magnitude of the 
CP violation effect for the $\nu_\mu\to\nu_\tau$ channel could attain 
as large as $|\Delta P| \simeq 0.22$ in the case of $\Delta m^2 \simeq 
3.5\times 10^{-3}$ ${\rm eV}^2$. Incidentally, we display the probabilities 
$P(\nu_\mu\to\nu_e)$ and $P(\bar{\nu_\mu}\to\bar{\nu_e})$ in Fig.9 
and $\Delta P(\nu_\mu\to\nu_e)$ by a dashed curve in Fig. 8. We show 
in Table 3 the $\Delta m^2/E$-dependence of $P(\nu_\mu\to\nu_e), 
P(\nu_e\to\nu_\tau), P(\nu_\mu\to\nu_\tau), \Delta P(\nu_\mu\to\nu_e), 
\Delta P(\nu_e\to\nu_\tau)$ and $\Delta P(\nu_\mu\to\nu_\tau)$ 
for $L = 730$ km. 

\begin{table}
\caption{$\Delta m^2/E$-dependence of the oscillation probabilities 
$P_{\alpha\beta}$ and their differences $\Delta P_{\alpha\beta}$ for the 
long-baseline experiment of $L = 730$ km. $P_{\alpha\beta}\equiv 
P(\nu_{\alpha}\to\nu_{\beta})$ and $\Delta P_{\alpha\beta}\equiv  
P(\nu_{\alpha}\to\nu_{\beta})-P(\bar{\nu_{\alpha}}\to
\bar{\nu_{\beta}})$. $\Delta m^2/E$ is in $10^{-3}$ ${\rm eV}^2/
{\rm GeV}$.}
\begin{center}
\setlength{\tabcolsep}{12pt}
\begin{tabular}{rrrrrrr} \hline
\multicolumn{1}{c}{$\Delta m^2/E$} & 
\multicolumn{1}{c}{$P_{\mu e}$} & 
\multicolumn{1}{c}{$P_{e\tau}$} & 
\multicolumn{1}{c}{$P_{\mu\tau}$} & 
\multicolumn{1}{c}{$\Delta P_{\mu e}$} & 
\multicolumn{1}{c}{$\Delta P_{e\tau}$} & 
\multicolumn{1}{c}{$\Delta P_{\mu\tau}$}   \\ \hline 
0.1 & 0.023 & 0.007 & 0.068 & 0.004 & 0.002 & -0.051  \\
0.2 & 0.025 & 0.008 & 0.053 & 0.008 & 0.004 & -0.101  \\
0.3 & 0.027 & 0.009 & 0.046 & 0.011 & 0.006 & -0.147  \\
0.4 & 0.028 & 0.010 & 0.047 & 0.014 & 0.007 & -0.188  \\
0.5 & 0.030 & 0.011 & 0.056 & 0.017 & 0.009 & -0.223  \\
0.6 & 0.031 & 0.012 & 0.072 & 0.019 & 0.010 & -0.251  \\
0.7 & 0.032 & 0.013 & 0.096 & 0.021 & 0.010 & -0.270  \\
0.8 & 0.032 & 0.014 & 0.126 & 0.021 & 0.011 & -0.281  \\
0.9 & 0.032 & 0.015 & 0.161 & 0.022 & 0.011 & -0.282  \\
1.0 & 0.032 & 0.016 & 0.201 & 0.021 & 0.010 & -0.274  \\
1.2 & 0.030 & 0.016 & 0.287 & 0.018 & 0.009 & -0.231  \\
1.4 & 0.028 & 0.016 & 0.374 & 0.012 & 0.006 & -0.158  \\
1.6 & 0.024 & 0.014 & 0.450 & 0.005 & 0.002 & -0.064  \\
1.8 & 0.020 & 0.012 & 0.504 & -0.003 & -0.001 & -0.038  \\
2.0 & 0.016 & 0.010 & 0.530 & -0.010 & -0.005 & 0.135  \\
2.5 & 0.011 & 0.005 & 0.460 & -0.021 & -0.011 & 0.279  \\  \hline
\end{tabular}
\end{center}
\label{tab3}
\end{table}
\vskip 0.1truecm

\vskip 0.7truecm

\centerline{\large\bf V  Conclusion}
\vskip 0.2truecm

We have derived the constraints on the neutrino mixing matrix by using 
the data from the solar neutrino deficit, atmospheric neutrino anomaly, 
LSND oscillation experiments, Bugey experiment and the CHORUS 
and NOMAD experiments along with the other accelerator and reactor 
experiments in the four-neutrino model with mass scheme of the two 
nearly degenerate pairs separated with the order of 1 eV. 
We have used the most general parametrization of the mixing matrix with 
six mixng angles and six phases applicable to both Majorana and Dirac 
neutrinos and have obtained the two serious constraints about the four of 
the six mixing angles and no constraints on the phases. 

By using these constraints, we have calculated the oscillation probabilities of 
$P(\nu_\mu\to\nu_e), P(\nu_e\to\nu_\tau)$ and $P(\nu_\mu\to\nu_\tau)$ 
and have investigated CP violation in the long-baseline neutrino oscillations of 
$\Delta P(\nu_\mu\to\nu_e), \Delta P(\nu_e\to\nu_\tau)$ and 
$\Delta P(\nu_\mu\to\nu_\tau)$. The quantity 
$\Delta P(\nu_\mu\to\nu_\tau)$ is found to be able to attain a value 
as large as $\pm 0.28$ due to the large mixing between 
$\nu_\mu$ and $\nu_\tau$ and the mass scheme of the four neutrinos and, 
therefore, it could be observed in the long-baseline experiments. 
We have shown that 
the contribution to $\Delta P(\nu_\mu\to\nu_\tau)$ from the matter effect is 
at most 0.01, 0.005 and 0.001 in magnitude for $\Delta m^2=1.0\times 
10^{-3}, 2.0\times 10^{-3}$ and $5.0\times 10^{-3}{\rm eV}^2$, 
respectively. So,we can conclude that the matter effect is negligibly small 
in comparison with the possible maximum value of the genuine CP-violation 
effect of $|\Delta P(\nu_\mu\to\nu_\tau)|=0.28$.

\newpage
\centerline{\bf Figure captions}

\vskip 1.0truecm
\noindent
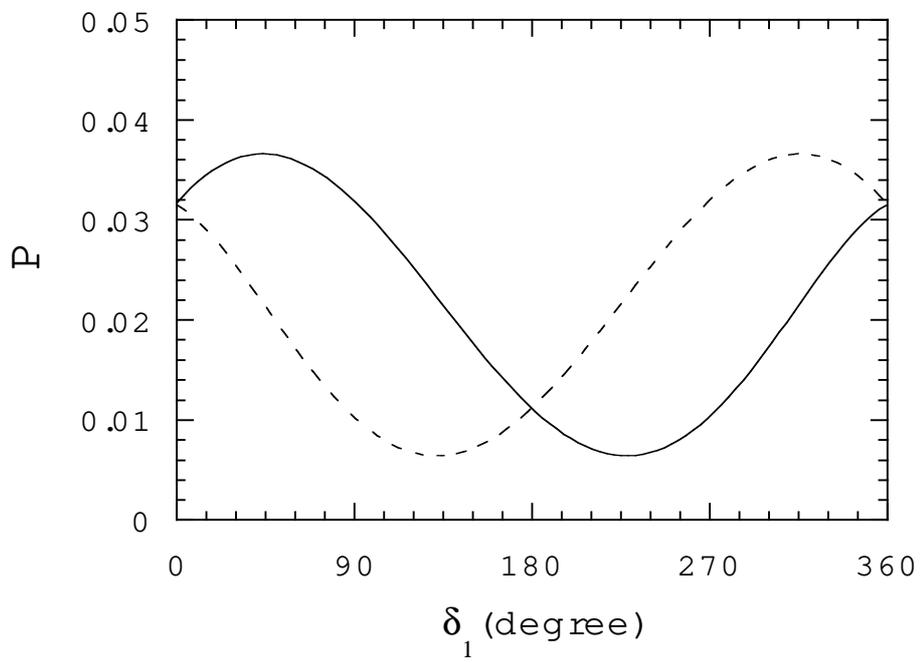
{\bf Fig.1.} The oscillation probability of $\nu_{\mu}\to\nu_e$(solid curve) 
and $\bar{\nu_{\mu}}\to\bar{\nu_e}$(dashed curve) with respect to the phase 
$\delta_1$ of the mixing matrix for the long-baseline experiment. The other angles 
and phases are fixed as $s_{02}=s_{03}=0.11 (c_{02}=c_{03}=0.994), 
s_{12}=0.91(c_{12}=0.415), s_{13}=0.67 (c_{13}=0.742), s_{01}=s_{23}
=1/\sqrt{2} (c_{01}=c_{23}=1/\sqrt{2}), \delta_{01}=\delta_{02}
=\delta_{03}=\delta_{12}=0$ and $\delta_2=\pi/2$. 
\vskip 0.5truecm

\noindent
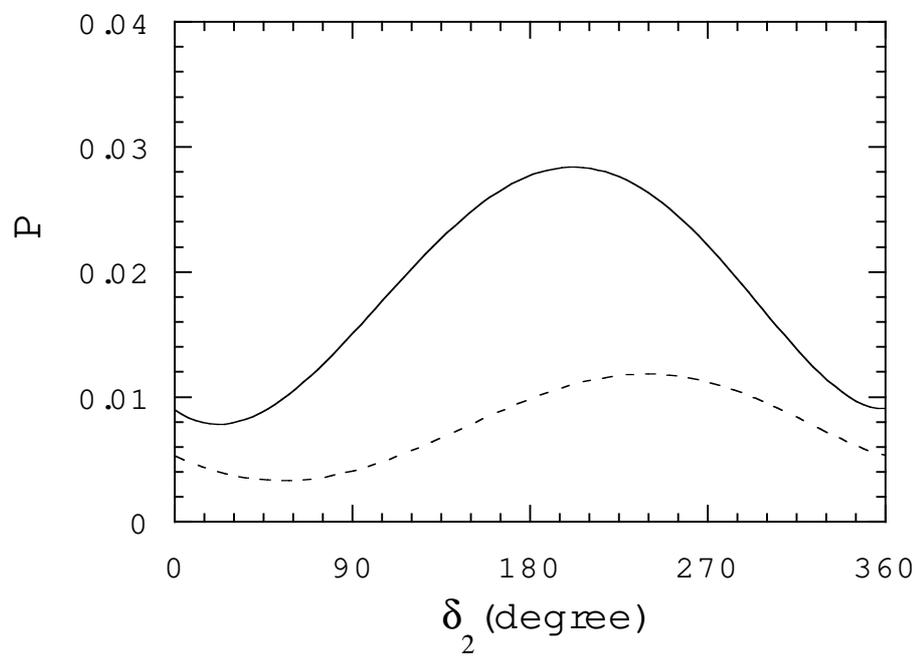
{\bf Fig.2.} The oscillation probability of $\nu_e\to\nu_{\tau}$(solid curve) 
and $\bar{\nu_e}\to\bar{\nu_{\tau}}$(dashed curve) with respect to the phase 
$\delta_2$ of the mixing matrix for the long-baseline experiment. The other angles 
and phases are the same as in Fig.1 except for $\delta_1=\pi/2$ fixed. 
\vskip 0.5truecm

\noindent
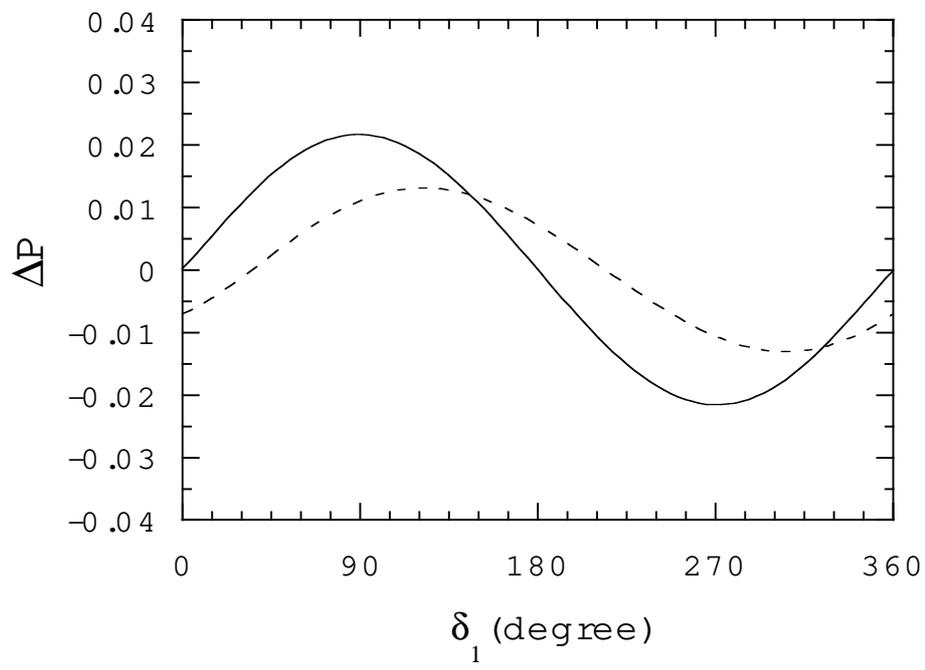
{\bf Fig.3.} The probability difference $\Delta P(\nu_{\mu}\to\nu_e)$ 
(solid curve) and $\Delta P(\nu_e\to\nu_{\tau}) $(dashed curve) with respect 
to the phase $\delta_1$ for the long-baseline experiment. The other angles 
and phases are the same as in Fig.1.
\vskip 0.5truecm

\noindent
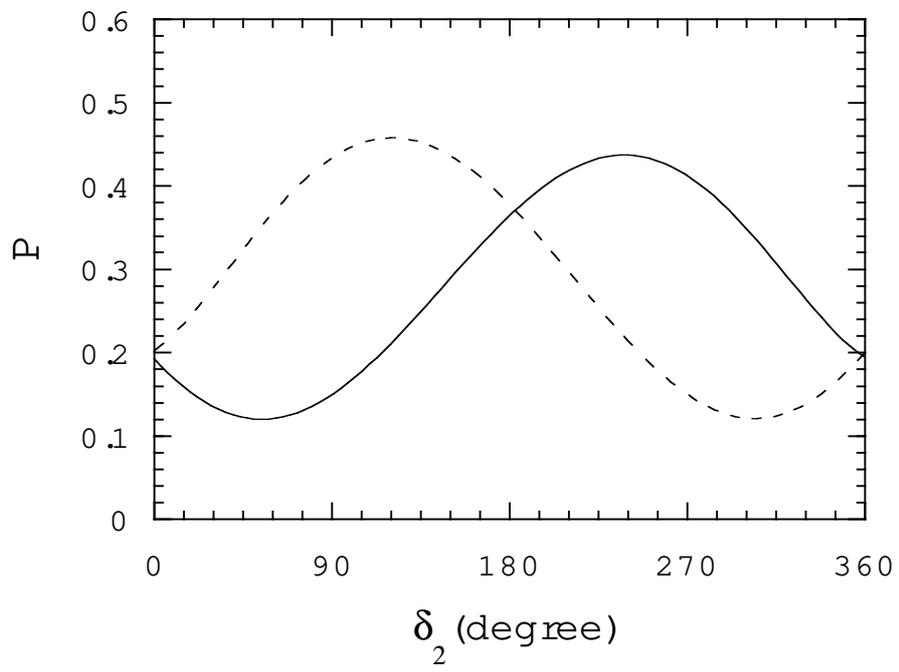
{\bf Fig.4.} The oscillation probability of $\nu_{\mu}\to\nu_{\tau}$
(solid curve) and $\bar{\nu_{\mu}}\to\bar{\nu_{\tau}}$(dashed curve) 
with respect to the phase $\delta_2$ for the long-baseline experiment. The other 
angles and phases are the same as in Fig.2. 
\vskip 0.5truecm

\noindent
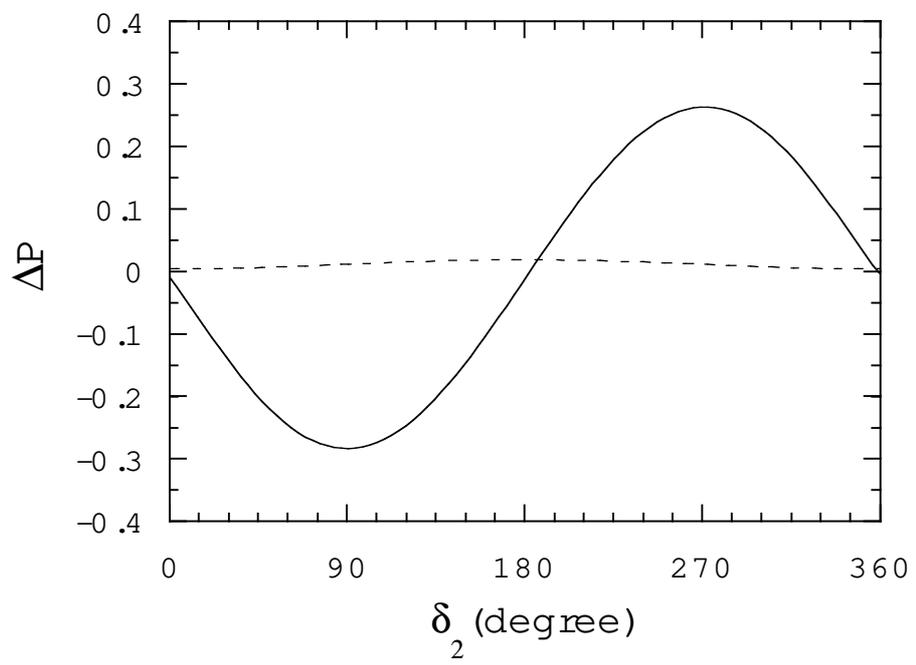
{\bf Fig.5.} The probability difference $\Delta P(\nu_{\mu}\to\nu_{\tau})$ 
(solid curve) and $\Delta P(\nu_e\to\nu_{\tau})$ (dashed curve) with respect 
to the phase $\delta_2$ for the long-baseline experiment. The other angles 
and phases are the same as in Fig.2.
\vskip 0.5truecm

\noindent
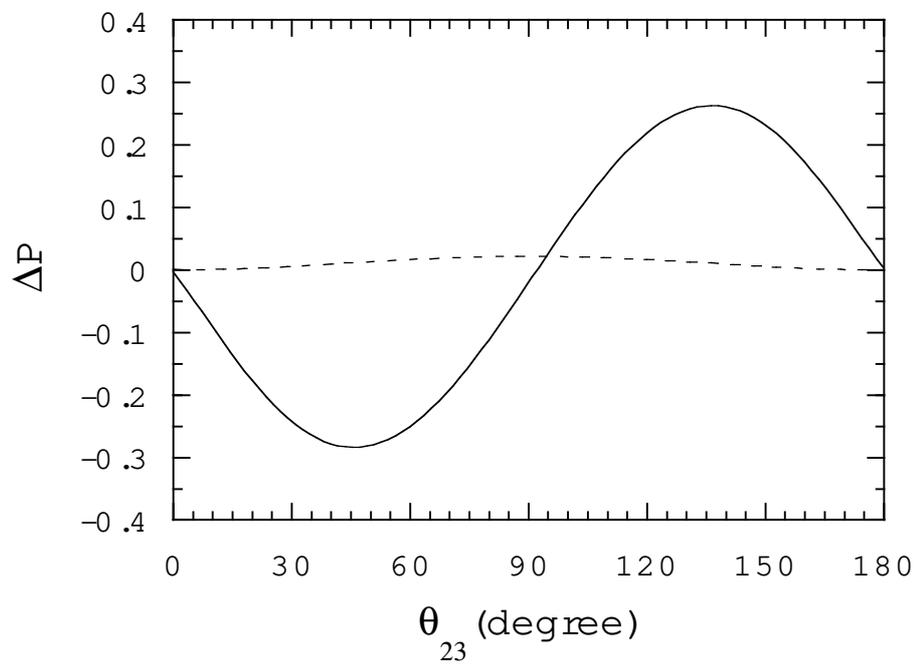
{\bf Fig.6.} The probability difference $\Delta P(\nu_{\mu}\to\nu_{\tau})$ 
(solid curve) and $\Delta P(\nu_e\to\nu_{\tau})$ (dashed curve) with respect 
to the angle $\theta_{23}$ for the long-baseline experiment. The other angles 
and phases are the same as in Fig.1 except for $\delta_1=\pi/2$ fixed.
\vskip 0.5truecm

\noindent
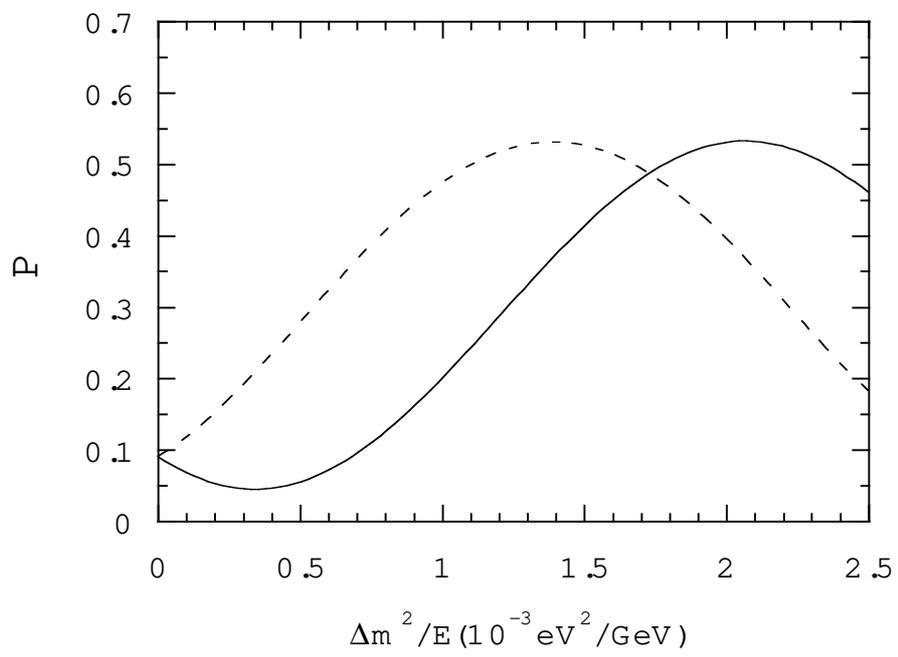
{\bf Fig.7.} The oscillation probability of $\nu_{\mu}\to\nu_{\tau}$
(solid curve) and $\bar{\nu_{\mu}}\to\bar{\nu_{\tau}}$(dashed curve) 
with respect to $\Delta m^2/E$ for the long-baseline experiment with the 
distance of $L=730$ km. The angles and phases are the same as in Fig.1 
except for $\delta_1=\pi/2$ fixed. 
\vskip 0.5truecm

\noindent
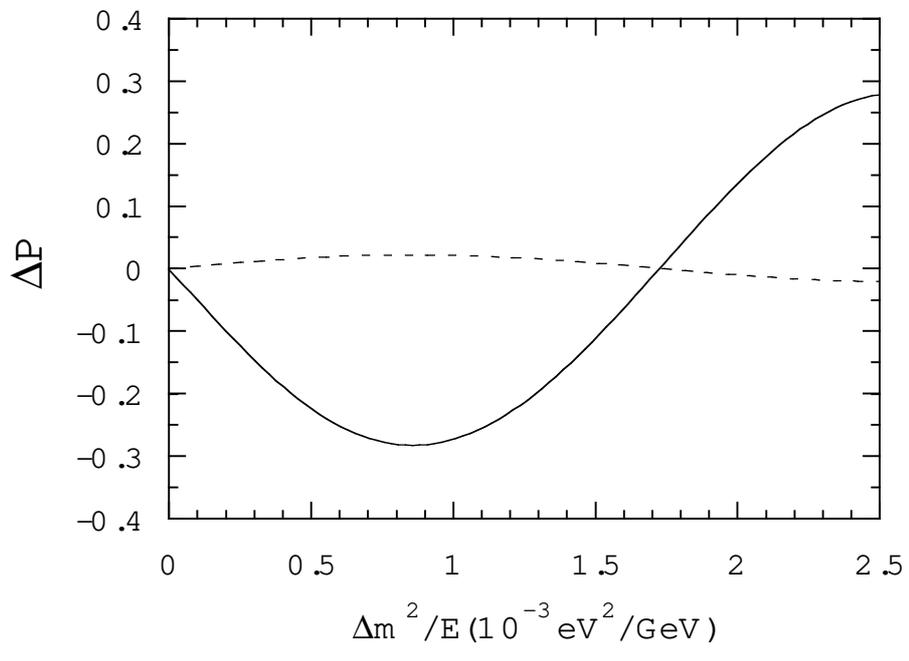
{\bf Fig.8.} The probability difference $\Delta P(\nu_{\mu}\to\nu_{\tau})$
(solid curve) and $\Delta P(\nu_{\mu}\to\nu_e)$(dashed curve) 
with respect to $\Delta m^2/E$ for the long-baseline experiment with the 
distance of $L=730$ km. The angles and phases are the same as in Fig.7. 
\vskip 0.5truecm

\noindent
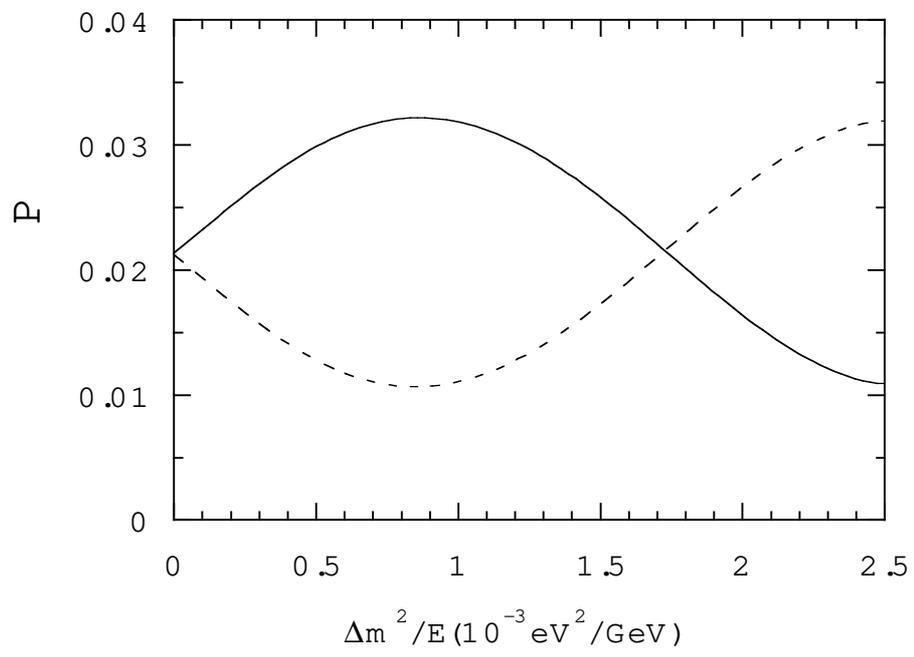
{\bf Fig.9.} The oscillation probability of $\nu_{\mu}\to\nu_e$(solid curve) 
and $\bar{\nu_{\mu}}\to\bar{\nu_e}$(dashed curve) with respect to 
$\Delta m^2/E$ for the long-baseline experiment with the distance of 
$L=730$ km. The angles and phases are the same as in Fig.7.
\vskip 0.5truecm

\end{document}